\documentclass[11pt,a4paper]{article}
\usepackage[utf8]{inputenc}
\usepackage[T1,T5]{fontenc}
\DeclareUnicodeCharacter{00FC}{{\fontencoding{T1}\selectfont\"u}}
\usepackage{lmodern}
\usepackage{microtype}
\usepackage[margin=2.5cm]{geometry}
\usepackage{setspace}
\usepackage{amsmath,amssymb}
\usepackage{graphicx}
\usepackage{booktabs}
\usepackage{array}
\usepackage{indentfirst}
\usepackage{placeins}

\usepackage{caption}
\usepackage[hyphens]{url}
\usepackage{xcolor}
\usepackage[colorlinks=true,urlcolor=blue!60!black,linkcolor=black,citecolor=black]{hyperref}
\title{An Event-Driven System Architecture for a 15-Minute Urban Flood Digital Twin: A Hanoi Case Study}

\author{Mai Anh Tran, Vy Nguyen, and Vova Nguyen\\[0.5em] \normalsize TECOTEC Technologies Joint Stock Company\\ \normalsize maianh@tecotec.com.vn, vynguyen@tecotec.tech}

\date{}

\begin{document}

\maketitle

\begin{abstract}

Flash floods in Hanoi submerge roads within minutes of heavy rain onset, yet public flood advisories remain retrospective. Hydrodynamic models could supply forward-looking depth guidance, but they are typically run offline for historical events, and prior studies have not evaluated whether physics-based simulation, spatial serving, and vehicle-aware routing can operate together within a fixed refresh cycle using a cloud solver and a single commodity server. This paper evaluates an end-to-end prototype of an automated flood digital twin designed to meet a 15-minute refresh cycle under data scarcity. The architecture organizes execution into decoupled, event-driven workflows that chain weather-gated 3Di simulations through hydraulic restart states, serve flood depths through a dual-track spatial layer, and publish road-edge depths via standardized output contracts. Evaluated on a 57-run storm replay in the Mỹ Đình and Tây Mỗ sector of Hanoi driven by satellite rainfall estimates, every cycle completed within the 15-minute deadline, publishing a ready edge table at a median of 537~s after slot start. Indexed raster tiles resolved point-depth lookups in a median of 0.54~ms (277 times faster than un-tiled rasters), while pre-rendered vector tiles served web maps without database queries. These findings demonstrate that a commodity server coupled to a cloud solver can sustain the 15-minute operational cycle and decouple routing policies from simulation internals. Because simulated depths were not validated against field gauges and the replay used observed satellite rainfall, the results establish system feasibility rather than forecast accuracy or route safety.

\end{abstract}


\section{Introduction}

\subsection{Flash Flooding and the Information Deficit}

Rapid urban expansion has increased the exposure of Southeast Asian cities towards extreme rainfall, since convective storms produce intense rain over short durations onto catchments where permeable land has been replaced by paved surfaces. Across Hanoi, convective downpours frequently overwhelm local collector networks and trigger rapid road inundation shortly after rain begins (Luo et al., 2018), while peripheral development between 2014 and 2024 has expanded high-risk flood zones faster than drainage capacity (Nguyen et al., 2026).

Under these conditions, forward-looking flood guidance is needed across pump control, emergency response, freight logistics, and daily commuting, which serves here as the demonstration case because whether a vehicle can pass a flooded street gives a concrete test of forecast usability. In Hanoi, however, public flood advisories remain retrospective. Hanoi Sewerage and Drainage Company (HSDC) reports flooded locations only after streets are already submerged, and neither these bulletins nor press coverage of recent storms provide a forward horizon, a predicted depth, or vehicle-specific guidance (HSDC, 2024; Kefi et al., 2018; Tuổi Trẻ, 2025). As a result, travelers must rely on personal judgment and fragmented social media reports, leaving a gap between the speed at which streets flood and the speed at which actionable information reaches road users.

\subsection{Limits of Conventional Modeling}

Conventional 2D hydrodynamic modeling would appear to be the natural remedy, and subgrid formulations resolve street-scale flow paths without a uniformly fine computational mesh (Stelling, 2012; van Dam et al., 2012). Even so, these models are typically run offline for historical events or planning scenarios, producing static hazard maps that cannot follow a moving storm front. Apel et al. (2024) showed that a flood warning workflow is useful only when a simulation finishes well inside the forecast update interval, excluding workflows whose runtime is measured in hours. However, a fast solver alone does not make a pipeline operational, because placing weather ingestion, numerical simulation, map rendering, and route calculation inside one tightly coupled script allows a delay in any downstream stage to stall the next simulation cycle (Caíno-Lores et al., 2020).

\subsection{The Digital Twin Paradigm and the 15-Minute Cycle}

Digital twins address both limits by coupling live meteorological input, numerical simulation, and downstream services into a continuously running system. Mandal et al. (2026) described geospatial twins for situational awareness during urban floods, noting that simulation frequency is often decoupled from real-time web delivery, while Kaynak et al. (2025) demonstrated that a city-scale twin remains operational only when weather inputs are linked automatically to infrastructure impacts. Taken together, these studies establish that the individual components exist, yet none evaluates whether physics-based 2D simulation, dual-track spatial serving, and network-edge depth publishing can operate together within a fixed refresh cycle using a cloud solver and a single commodity server.

Closing this gap requires a refresh cycle matched to how fast Hanoi streets flood. Press reports of recent storms indicate that 30 minutes of heavy rain is sufficient to submerge major streets (Tuổi Trẻ, 2026). Because a 30-minute refresh cycle could leave guidance as old as the flood onset itself, this system adopts a 15-minute refresh cycle so that at least two updates are issued before inundation peaks. Each 15-minute cycle computes a 2-hour (7,200~s) forward-looking forecast horizon across eight 15-minute steps, ensuring that predicted depths for the upcoming 30 minutes are already stored when a consumer queries the system.

\subsection{Objective and Contributions}

This paper presents the design, implementation, and replay evaluation of an end-to-end prototype of an automated flood digital twin for Hanoi that turns rainfall inputs into road-edge depth tables within a 15-minute cycle under data scarcity (Bui \& Tran, 2026). Four contributions follow from this objective. Firstly, the pipeline is organized into decoupled, event-driven workflows that isolate faults so that a downstream failure does not stall the hydraulic chain, as verified under an injected routing failure. Secondly, weather-gated simulation runs are chained through hydraulic restart states, keeping simulated floodwaters continuous across forecast cycles. Thirdly, a dual-track spatial serving layer supports both sub-millisecond point queries and static web map tiles without loading the simulation path. Last but not least, standardized output contracts expose road-edge depths through stable network identifiers, decoupling the simulation pipeline from downstream routing policies.

\section{Theoretical Background and Related Work}

\subsection{Workflow Orchestration in Scientific Computing}

Environmental forecasting workflows written as monolithic batch scripts suffer from input-output serialization, storage growth, and cascading stalls when secondary formatting or indexing tasks fail (Caíno-Lores et al., 2020). To isolate stages, workflow engines represent pipelines as directed acyclic graphs (DAGs) with task-level state tracking and retry rules. Nagpal (2026) benchmarked modern orchestrators and reported that Apache Airflow provides dependable metadata-driven scheduling for multi-stage pipelines. However, both studies evaluate batch workflows without a strict turnaround deadline, leaving open whether asset-driven DAG decoupling can sustain a chained hydrodynamic simulation inside a 15-minute service level agreement.

\subsection{Hydrodynamic Simulation}

Urban flood propagation is governed by the 2D shallow water equations, where flow paths are controlled by kerbs, embankments, and buildings much smaller than a practical grid cell. Stelling (2012) proposed the subgrid quadtree method, in which a coarse computational mesh integrates volume and conveyance over a high-resolution terrain model, and van Dam et al. (2012) implemented this approach in the 3Di solver to achieve event-scale simulation within minutes. Because flash floods transition rapidly between dry ground and thin water films, Casulli and Stelling (2011) further showed that semi-implicit integration conserves mass and remains unconditionally stable across wetting and drying fronts, a property essential for unattended operation. For operational warning, Apel et al. (2024) argued that runs must finish well inside the forecast update interval and avoid the spin-up distortion of starting cold from a dry surface, yet prior work does not report how a warm-started subgrid chain behaves across a full storm on commodity hardware.

\subsection{Spatial Data Storage and Multi-Scale Serving}

A flood model produces a dense depth raster at every forecast step, which point-level risk analysis and interactive web maps consume in incompatible ways (Xie et al., 2026). Storing a full raster as a single database row forces the query engine to decode the entire grid for a single coordinate lookup, whereas splitting the raster into small GiST-indexed blocks lowers point-query latency by orders of magnitude (Agosto, 2013) and supports spatio-temporal archiving as runs accumulate (Zimányi et al., 2020). For web visualization, Web Mercator tiling restricts client downloads to the active viewport (Stefanakis, 2017), adaptive vector tile pyramids bound browser memory on dense layers (Li \& Li, 2024), and object storage decouples tile generation from client request load (Wan et al., 2016). However, each of these studies addresses a single access pattern rather than serving both indexed point lookups and static vector tiles from a solver that publishes a new surface every 15 minutes.

\subsection{Digital Twins and Interface Contracts}

A flood digital twin couples physical observations or forecasts to a numerical model through automated updates (Kaynak et al., 2025; Ruangpan et al., 2023). Once data are flowing, Mandal et al. (2026) concluded that synchronizing a multi-minute hydraulic solver with downstream decision services that expect sub-second responses remains a central architectural bottleneck. Moreover, existing platforms rarely specify a formal data contract between the solver and its consumers, meaning that direct queries inherit solver latency and schema changes can break downstream adapters. Compared with this missing output contract, input gating is well established, as Kim et al. (2025) demonstrated that quality filters and rainfall thresholds prevent forecast noise from triggering expensive simulations; hence, the present study adopts threshold gating and focuses on the orchestration, serving, and contract architecture.

\section{Description of the Case Study and Available Data}

\subsection{Study Area and Flood Exposure}

The case study covers a 32.8 km\textsuperscript{2} urban sector in western Hanoi around Mỹ Đình and Tây Mỗ, extending 9.9~km east to west and 5.7~km north to south (see Figure 1). The Nhuệ River and retention lakes such as Mễ Trì and Miếu Đầm cross the eastern sector, while Thăng Long Boulevard carries arterial traffic through depressed underpasses that collect runoff during heavy rain (Bui \& Tran, 2026). Across Hanoi, Nguyen et al. (2026) found that high-risk flood zones cover roughly 25\% of the municipal area following rapid urban expansion between 2014 and 2024. Luo et al. (2018) demonstrated that rainfall intensities above 50~mm/h overwhelm Hanoi's open collector canals and submerge major intersections under 0.2 to 0.5~m of water within 15 to 30 minutes, while Kefi et al. (2018) concluded that transport and property damages in the adjacent Tô Lịch watershed are compounded by local drainage deficits and the lack of real-time warning.

\begin{figure}[!htbp]
\centering
\includegraphics[width=0.92\linewidth]{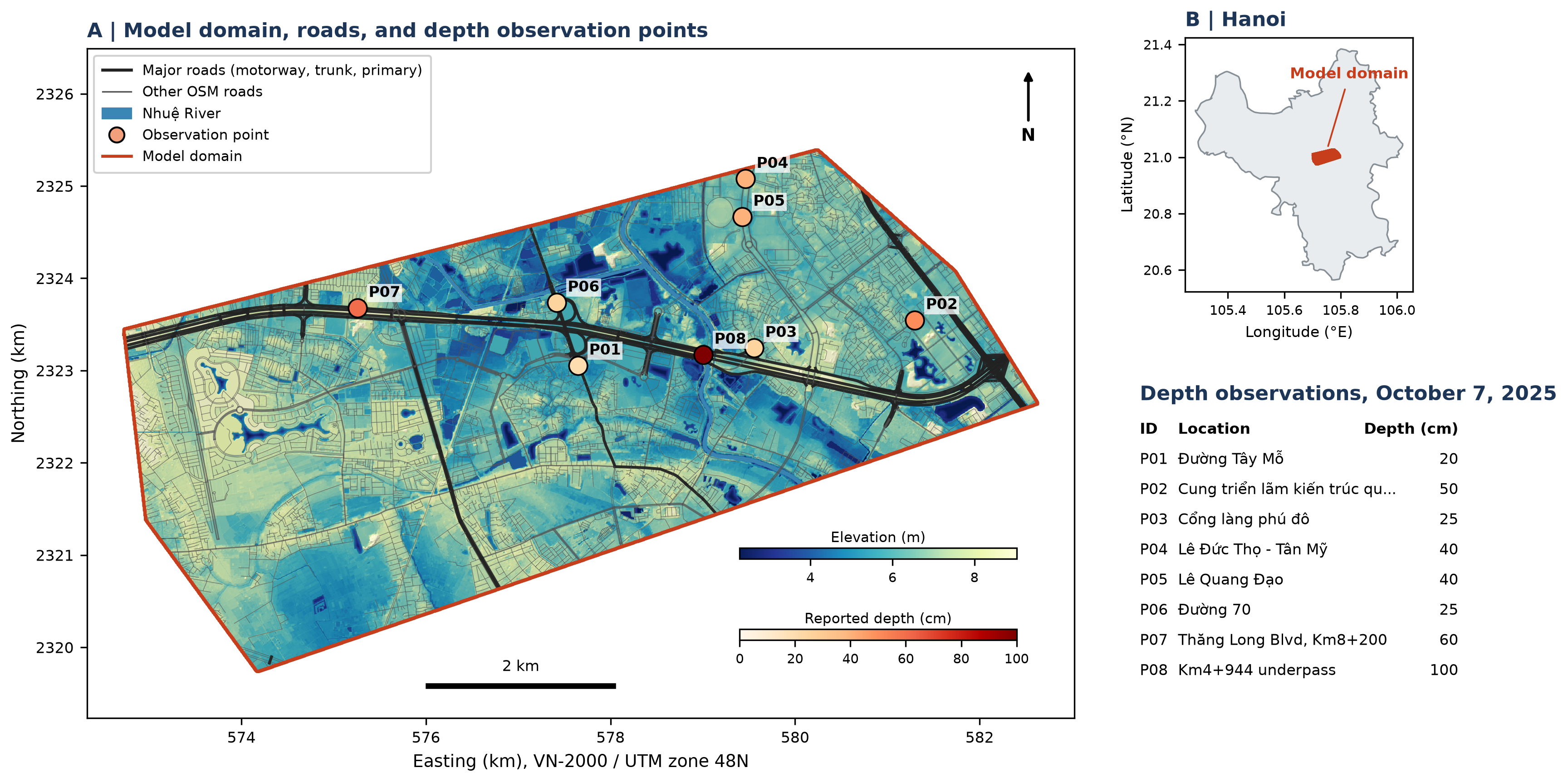}
\caption*{\textbf{Figure 1}\\[0.2em]\emph{Study Area}\\[0.5em]{\small \emph{Note.} (A) Model domain, outlined as the area of cells that hold elevation, on the 1~m digital elevation model, with OpenStreetMap roads, the Nhuệ River, and the eight points where news reports of October 7, 2025 give a flood depth. (B) Location of the domain in Hanoi.}}
\end{figure}

\subsection{Flood Reporting and the Informational Deficit}

Public reporting during the October 7, 2025 flood following Typhoon Matmo illustrates the informational delay facing Hanoi commuters. Tuổi Trẻ (2025) published its live flood report at 07:10, more than seven hours after heavy rain began at 00:00 and over an hour after water submerged Thăng Long Boulevard at 05:30, describing depths only qualitatively (e.g., water up to a pedestrian's waist). Similarly, Tiền Phong (2025a) published the Hanoi Department of Construction's official list of more than 100 flooded locations, including 23 impassable sites with depth ranges, at 19:28 that evening, long after morning commuters had already entered flooded corridors.

\subsection{Rainfall Forcing Data}

To provide forward-looking forcing before water accumulates, the pipeline ingests 15-minute precipitation forecasts across a two-hour horizon from the Tomorrow.io weather service. For reproducible evaluation, this study replays the convective storm of September 16 and 17, 2026 using half-hourly satellite precipitation estimates from the NASA GPM IMERG Late Run product on a 0.1$^{\circ}$ grid (Huffman et al., 2023), area-weighted across the model domain. The full 48.5-hour series totals 84.95~mm with a peak 30-minute mean rate of 8.73~mm/h; the replay uses the wettest 12 hours (53.88~mm between 23:00 on September 16 and 11:00 on September 17, +07:00), which is gentler than the unwindowed 177.3 to 199~mm station totals reported by the municipal drainage operator for that night (Nhân Dân, 2026).

During the replay, the pipeline slices a nine-step (two-hour) forward window at each 15-minute slot without temporal interpolation, shifting the 12-hour storm window to run in wall-clock time from 04:45 to 16:45.

\subsection{The Hydrodynamic Model and Its Limits}

The hydraulic core is the 3Di subgrid model of Mỹ Đình and Tây Mỗ developed by Bui and Tran (2026), which was the only operational 2D model available for this sector of Hanoi. Because the model contains 46 levees and a simple infiltration loss over the 1~m terrain grid but lacks surveyed storm sewers, pumping stations, boundary conditions, or ground rain gauges, rainfall is the sole inflow and infiltration the sole outflow, meaning that unmodeled pump operations such as the Yên Nghĩa station running at 60 to 70\% capacity (Tiền Phong, 2025b) are not represented. On the October 7, 2025 flood, the model reproduced reported depths within 2.2~cm at six of the eight calibration points in Figure 1 (mean absolute error 1.34~cm) while remaining dry at Tây Mỗ Road and Road 70 (overall mean absolute error 6.63~cm; Bui \& Tran, 2026); consequently, simulated depths serve as hazard indicators for defensive gating rather than deterministic ground truth.

\subsection{Road Network Data}

The road network is derived from OpenStreetMap vector geometries clipped to the model domain and filtered to vehicular classes (e.g., motorways, trunk roads, arterials, avenues, connectors, and residential streets), yielding 67,833 discretized road edges. Because standard routing parsers reassign sequential integer identifiers whenever a map extract is re-imported, each edge is assigned a stable composite key combining the 64-bit OpenStreetMap way identifier and the zero-indexed segment position along the way. Re-importing the same extract preserved all 67,833 keys identically, whereas comparing an older and a newer extract showed that 2.3\% of keys on ways with unchanged edge counts (1,446 of 62,936) shifted geometry and 752 ways changed edge count, requiring a geometry check before carrying depth tables across map extracts.

\section{Methods and System Architecture}

\subsection{Weather Ingestion, Gating, and Stateful Simulation Chaining}

The pipeline runs on a single commodity server with an Intel Core i5-10400F processor (6 cores, 12 threads) and 31.9~GB of memory, where the Docker virtual machine shares 12 logical CPUs and 20~GB of memory across containers. The 3Di solver runs in the 3Di cloud service, which the worker starts and polls through its programming interface before downloading the results file so that depth extraction, tile building, and edge mapping execute on the server without manual intervention between cycles. Each cycle retrieves quantitative precipitation forecasts from the Tomorrow.io weather service at 15-minute intervals across a two-hour forward horizon (7,200~s), giving an eight-step rainfall sequence as the forcing of the simulation.

Given that a full 2D simulation in dry weather wastes capacity, and that automated threshold gates are established practice for stormwater forecasting platforms (Kim et al., 2025), a gate is applied before any simulation is scheduled. A chain of simulations starts from a dry state only when the forecast intensity within the lookahead window reaches the operational rainfall threshold $R_{\text{thresh}} = 5.0\text{ mm/h}$, which is roughly ten times below the 50~mm/h at which Luo et al. (2018) found that the open collector canals of Hanoi are overwhelmed. However, surface water frequently lingers along low-lying corridors after the rain ends, so stopping the chain when the rain stops would end the simulation while roads are still flooded. Therefore, the pipeline applies a standing-flood continuation rule, which runs the next 15-minute slot $t$ of an active chain whenever rain is still forecast or water still stands on the road network:

\begin{equation}
\Omega(t) = \Gamma_{\text{rain}}(t) \lor \Gamma_{\text{flood}}(t).
\end{equation}

Once a chain has started, $\Gamma_{\text{rain}}(t) = 1$ whenever any rain is forecast, while $\Gamma_{\text{flood}}(t)$ reads the deepest water depth $d_e(t)$ that each road edge $e \in E$ reaches over all steps of the run at slot $t$, excluding ponds and rivers:

\begin{equation}
\Gamma_{\text{flood}}(t) = \begin{cases} 1 & \text{if } \max\limits_{e \in E,\; k \in \{1,\dots,K\}} d_e(t-k) \ge h_{\text{thresh}}, \\ 0 & \text{otherwise,} \end{cases}
\end{equation}

with a depth threshold $h_{\text{thresh}} = 5\text{ cm}$ and a window of $K = 5$ consecutive runs (75 minutes of refresh time). The 5~cm threshold reflects the water depth below which surface ponding ceases to obstruct vehicular or pedestrian movement, and requiring $K = 5$ consecutive runs ensures that no rainfall is predicted and all road depths remain below 5~cm for at least three hours across the cycles and their 2-hour forecast horizon before the pipeline returns to idle.

Starting each cycle from dry conditions would erase standing water on the roads and add the spin-up of a cold start, so successive cycles in an active chain are linked through a hydraulic warm start:

\begin{equation}
\mathbf{H}_t = \Phi\left(\mathbf{H}_{t-1}, P(t)\right),
\end{equation}

where $\Phi$ denotes the 3Di shallow water integration operator, $\mathbf{H}_t$ the hydraulic state vector, and $P(t)$ the precipitation forcing. Each run of the 3Di cloud service registers two saved states, one 900~s and one 1,800~s after the start of its slot. The state at 900~s is valid at the next cycle time, so the run of cycle $t$ starts from the state that cycle $t-1$ saved for that exact moment, while the state at 1,800~s serves as a backup up to two cycles (30 minutes) older. When no such state exists, or when no run of the previous slot succeeded or remained running, a run starts cold from the dry template. Each 15-minute cycle thus continues the physical state of the previous one while computing an eight-step forecast window, after which eight parallel worker processes extract the 2D maximum water depth grid of each interval from the local container disk into georeferenced rasters.

\subsection{Decoupled Multi-DAG Workflow Orchestration}

Once the rasters exist, the remaining stages must not be able to stall the simulation chain that produced them. Since tightly coupled scientific workflows suffer cascading task failures (Caíno-Lores et al., 2020), and a DAG scheduler provides dependency isolation across multi-stage pipelines (Nagpal, 2026), the digital twin is formalized as an asynchronous state transition system coordinated by data-aware asset triggers $\mathcal{E}$. As illustrated in Figure 2, execution is partitioned across four independent Directed Acyclic Graphs (DAGs) in Apache Airflow. The simulation DAG ($\mathcal{D}_{\text{sim}}$) covers weather ingestion, rainfall gating, the 3Di simulation, depth extraction, and raster tile loading. The vector tile DAG ($\mathcal{D}_{\text{tile}}$) handles polygon extraction, dissolving, and vector tile pyramid generation, whereas the routing adapter DAG ($\mathcal{D}_{\text{route}}$) samples depths along road edges and serializes the binary impedance table. Finally, the logging DAG ($\mathcal{D}_{\text{hist}}$) records time-series performance metrics and edge-depth records in a time-series database.

These four DAGs are joined solely by Airflow asset events rather than time-based polling or direct triggers. When $\mathcal{D}_{\text{sim}}$ finishes depth extraction, it triggers $\mathcal{D}_{\text{tile}}$ in parallel with the raster load, and once the raster tiles are loaded it emits a versioned asset event $\mathcal{E}_{\text{raster}}$ that activates $\mathcal{D}_{\text{route}}$. Within $\mathcal{D}_{\text{route}}$, the edge table task aggregates water depths along the road segments and emits the edge asset event $\mathcal{E}_{\text{edge}}$, which triggers $\mathcal{D}_{\text{hist}}$ without blocking any upstream worker. This event chain sets a strict fault boundary between computational tiers: only tasks of $\mathcal{D}_{\text{sim}}$ read and write the simulation state file, so if the routing adapter meets a timeout, database lock, or memory spike, $\mathcal{D}_{\text{sim}}$ keeps its successful state and preserves the hydraulic warm start for the next cycle, as confirmed by a targeted fault-injection test. In addition, five tasks of $\mathcal{D}_{\text{sim}}$ inspect the execution timestamp and skip any run whose scheduled slot is older than the 120-minute forecast horizon, preventing a delayed run from overwriting a fresh forecast.

\begin{figure}[!htbp]
\centering
\includegraphics[width=0.92\linewidth]{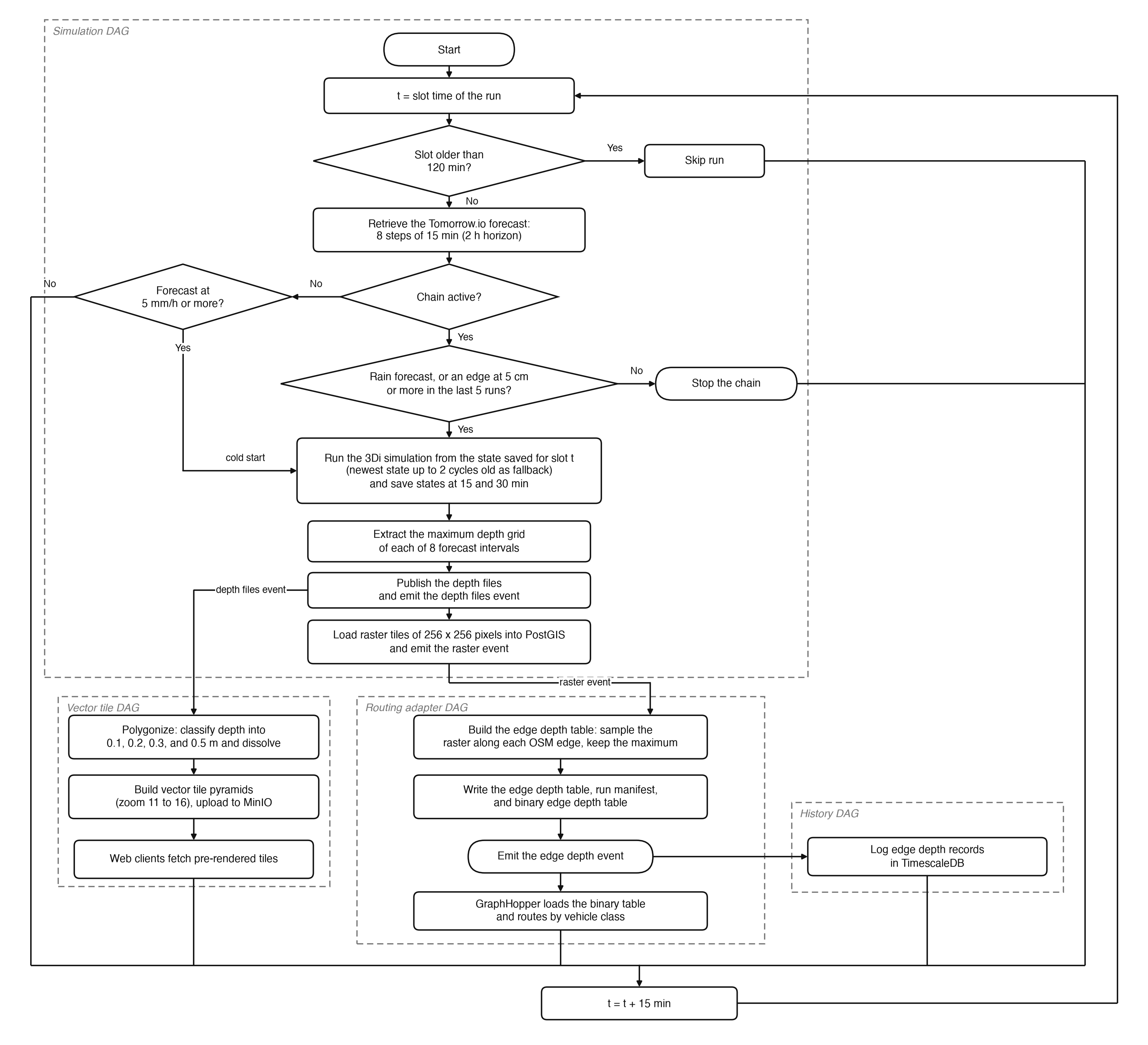}
\caption*{\textbf{Figure 2}\\[0.2em]\emph{Flowchart of One 15-Minute Cycle of the Flood Digital Twin in Its Later Layout, in Which the Tile Build Runs in Its Own Workflow}\\[0.5em]{\small \emph{Note.} Dashed frames mark the four Airflow DAGs, with $t$ as the slot time of the run. A chain of simulations starts at a forecast of 5~mm/h and continues for as long as rain is forecast or a road edge reached 5~cm in the last 5 runs. A run starts from the state saved for its slot or, failing that, from the newest state at most two cycles older, and it starts cold when no such state exists.}}
\end{figure}

\subsection{Dual-Track Spatio-Temporal Serving Engine}

Each forecast step produces a continuous depth field $D(x, y, t)$ on a 1~m grid of 10,077 by 5,436 cells (about 55 million cells and 209~MB once decoded). Sending monolithic rasters to browser clients congests the network and exhausts browser memory (Xie et al., 2026), so the serving engine separates point-level querying from multi-scale map visualization across two storage tracks. The first track stores the flood surface in PostGIS as uniform $256 \times 256$ pixel raster tiles covered by a GiST spatial index (Agosto, 2013):

\begin{equation}
\mathcal{T}_{i, j}(t) = \left\{ D(x, y, t) \mid x \in [x_i, x_i + \Delta X], \, y \in [y_j, y_j + \Delta Y] \right\},
\end{equation}

where $\Delta X, \Delta Y$ are the $256\text{ m} \times 256\text{ m}$ tile extents under the national coordinate reference system. Each time step splits into 880 tiles, and stating the bounding-box test explicitly in the query allows PostGIS to decode only the single tile that holds a queried coordinate.

The second track serves interactive web maps by combining Web Mercator viewport tiling (Stefanakis, 2017) with object storage that decouples tile generation from high-concurrency client requests (Wan et al., 2016). The tile DAG uses GDAL to classify depth rasters into inundation bands of 0.1~m, 0.2~m, 0.3~m, and 0.5~m, dissolves contiguous polygons with one process per band to remove tile-boundary artifacts, and compiles Mapbox Vector Tile pyramids in the $\{z\}/\{x\}/\{y\}.\text{pbf}$ layout across zoom levels 11 to 16 for delivery from a self-hosted MinIO instance. Client browsers thus fetch pre-rendered tiles through static HTTP requests that place zero query load on PostgreSQL.

To prevent storage exhaustion on commodity hardware while sustaining query throughput (Zimányi et al., 2020), data across the pipeline are organized into four storage tiers: PostGIS for the active $256 \times 256$ raster tiles, MinIO for pre-rendered vector tile pyramids, PostgreSQL relational tables for road-edge depth records and publication metadata, and TimescaleDB hypertables with chunk-level compression for historical telemetry and long-term depth logs. An automated cleaner prunes relational and raster records past a rolling retention window, while three forecast modes govern client display: live mode for active storm forecasts, pending mode before rainfall onset, and demonstration mode, which pins the peak flood frame of a historical run so dry-weather demonstrations require no cloud simulation.

\subsection{Core Output Contracts and Edge-Depth Mapping}

The digital twin exposes its outputs through two formal Core Output Contracts so external consumers never touch internal pipeline tables. Contract A governs spatial visual delivery through the $256 \times 256$ PostGIS raster tiles and the Mapbox Vector Tile pyramids. Contract B defines the tabular and routing interfaces, consisting of a relational table of road-edge depths, a single publication log table that records the run name, output type, and timestamp under read-only database roles, and a contiguous binary lookup file of 1,085,487 bytes (about 1.08~MB) that serializes the water depths of all 67,833 road edges across eight forecast steps for direct memory loading without database query latency.

Because dynamic environmental impedances can be evaluated over a static graph topology without structural modification (Bast et al., 2016), the road network retains its stable composite keys across runs. For each forecast step, the edge table task samples the depth raster along each edge geometry through a spatial overlay and assigns $d_e(t)$ as the maximum sampled depth rather than the mean, preserving localized depressions where ponding is deepest. On one step of one run with 4,600 wet edges, taking the all-sample mean would have reopened more than half of the edges that the maximum closes across every vehicle class. Contract B therefore carries only maximum edge depths and leaves all vehicle closure thresholds, precautionary margins, and routing cost penalties to the downstream routing consumer, which is evaluated in a companion study (Tran et al., 2026).

\section{Results}

\subsection{Pipeline Timing}

The storm replay executed 57 consecutive runs across the 15-minute slots from 04:00 to 18:00 (+07:00), with three dry-lead slots before rainfall onset at 04:45 and 54 storm slots. All 56 runs after the initial cold start recorded a hydraulic warm start, carrying the flood state continuously across the 14-hour replay without a break. The 5~mm/h rainfall gate passed 45 of the 57 slots directly, while the remaining 12 slots (four mid-storm lulls with forecast maxima of 3.6 to 4.5~mm/h and eight post-storm recession slots) continued automatically under the standing-flood rule. In seven separate dry-weather checks against the live forecast service prior to the replay, each check completed in about 2~s and suppressed the simulation as intended.

\begin{table}[!htbp]
\centering
\footnotesize
\setlength{\tabcolsep}{4pt}
\caption*{\textbf{Table 1}\\[0.2em]\emph{Task Timings of the Storm Replay on the Earlier Workflow Layout, All 57 Runs (Seconds)}}
\begin{tabular}{>{\raggedright\arraybackslash}p{0.27\textwidth}c c r r >{\raggedleft\arraybackslash}p{0.10\textwidth}>{\raggedleft\arraybackslash}p{0.10\textwidth}r}
\toprule
\textbf{Stage} & \textbf{Workflow} & \textbf{Runs} & \textbf{Mean} & \textbf{Median} & \textbf{95th percentile} & \textbf{99th percentile} & \textbf{Maximum} \\
\midrule
Solver step (start of the 3Di simulation and wait for its result) & Main & 57 & 121.0 & 133.3 & 144.6 & 144.9 & 145.1 \\
Result download & Main & 57 & 8.7 & 8.6 & 12.4 & 12.8 & 12.8 \\
Depth step (depth rasters from the solver output) & Main & 57 & 169.5 & 169.0 & 172.2 & 173.2 & 173.4 \\
Depth raster load into PostGIS & Main & 57 & 143.3 & 144.1 & 155.8 & 158.8 & 161.6 \\
Tile build & Main & 57 & 179.6 & 189.3 & 230.3 & 232.5 & 234.3 \\
Edge table build & Routing & 57 & 84.8 & 89.0 & 107.9 & 108.6 & 109.2 \\
Critical path (first task start to last task end) & Main & 57 & 491.5 & 488.0 & 542.2 & 633.8 & 734.0 \\
\bottomrule
\end{tabular}
\end{table}

Because the tile build and the depth raster load start in parallel after the depth step on the earlier workflow layout while the edge table runs in the routing workflow, the critical path of the main workflow is shorter than the sum of stage means. As shown in Table 1, all 57 runs finished within 734.0~s of the 900~s cycle budget, with a mean critical path of 491.5~s (median 488.0~s) and a maximum task-duration sum of 553~s. The depth extraction step was the most stable stage (95th percentile 172.2~s), indicating that its cost is independent of flooded area, whereas the vector tile build varied up to 234.3~s as inundated polygon complexity grew. Measured from the start of each slot, the median time to the first published map step was 456~s, and the median time to a ready edge table was 537~s (about 9 minutes), leaving 6 minutes of margin before the next cycle.

Initial pipeline runs prior to the replay exceeded the 15-minute budget (summing to about 17 minutes) because the eight parallel depth extraction workers read their inputs across a Windows host bind mount. In two controlled paired comparisons, copying the inputs to the container's local disk in 1.0~s reduced eight-step depth extraction from 368.6~s to 148.6~s in the first pair and from 349.6~s to 145.0~s (a 58.5\% reduction) in the second. Airflow daily medians across 17 operating days confirmed this reduction, dropping from 266-477~s before the change to 169-194~s during the replay period. Furthermore, five clean standing-flood slots on the later four-DAG layout showed that moving the vector tile build into its own workflow ($\mathcal{D}_{\text{tile}}$) alongside the raster load reduced the main-workflow critical path to 378-403~s, with tile builds taking 55-69~s and edge table builds taking 23-25~s.

\subsection{Spatial Serving and Storage}

On an idle server, point-depth lookups across 1,000 road-edge midpoints on the $256 \times 256$ GiST-indexed PostGIS tiles achieved a median latency of 0.54~ms (95th percentile 0.81~ms), compared to 149.5~ms (95th percentile 192.7~ms) on an un-tiled full-raster row of 10,077 by 5,436 pixels, a 277-fold speedup. Under active server load, tiled lookups maintained a median of 0.77~ms against 223.1~ms for the un-tiled row, and query times remained invariant as the raster table grew from one stored run to seven. For the map visualization track, fetching 380 pre-rendered vector tiles through the nginx proxy across 3,420 requests yielded median latencies of 3.1~ms for 1 client, 15.5~ms for 10 concurrent clients, and 67.3~ms for 50 concurrent clients (95th percentile 171.5~ms) with zero failed requests and zero rise in PostgreSQL transaction counts or rows returned above an idle control.

\subsection{Edge Depths and Contract Verification}

Building the Contract B road-edge depth table on a clean test database required 67.8 to 81.5~s across eight builds (mean 77.2~s). Comparing the first step of six selected runs against the depth forecast produced by the preceding run for that same timestamp confirms that hydraulic warm-start chaining preserves edge-level continuity across cycle boundaries, with wet-edge counts (depth $\ge 1\text{ cm}$) differing by at most 3 edges across all six transitions (Table 2). End-to-end interface verification with the downstream routing consumer (detailed in the companion routing study; Tran et al., 2026) further confirmed that loading the 1.08~MB Contract B binary table directly into memory kept time-dependent route queries within twice the dry-baseline latency across all test cells and produced zero crossings of roads closed by the published depth thresholds.

\begin{table}[!htbp]
\centering
\setlength{\tabcolsep}{10pt}
\caption*{\textbf{Table 2}\\[0.2em]\emph{Wet Edges in the First Step of Selected Runs, With the Forecast of the Previous Run for the Same Step}}
\begin{tabular}{l r r}
\toprule
\textbf{First step of the run (+07:00)} & \textbf{Wet edges} & \textbf{Forecast of the previous run} \\
\midrule
05:15 & 2,106 & 2,107 \\
07:00 & 3,513 & 3,510 \\
08:30 & 3,625 & 3,628 \\
11:00 & 4,067 & 4,067 \\
13:30 & 4,010 & 4,012 \\
16:00 & 4,483 & 4,481 \\
\bottomrule
\end{tabular}
\end{table}

\section{Discussion}

\subsection{Feasibility and System Tradeoffs on Commodity Hardware}

Completing all 57 storm replay runs within 734.0~s of the 900~s cycle demonstrates that a single commodity server paired with a cloud hydrodynamic solver can sustain an operational 15-minute update cadence without overlapping consecutive cycles, satisfying the real-time warning criterion of Apel et al. (2024). Because the 3Di solver executes in the cloud while raster extraction, spatial indexing, tile rendering, and edge mapping execute locally, this result establishes the feasibility of hosting the data-intensive post-processing and serving tiers on modest municipal hardware rather than running the full 2D shallow-water solver on one machine. Moreover, the 58.5\% reduction in depth extraction time achieved by staging inputs on the container's local disk rather than reading across a host bind mount shows that I/O path contention can dominate arithmetic cost in containerized geospatial workflows.

Partitioning the pipeline into four asset-triggered DAGs aligns with Caíno-Lores et al. (2020) and Nagpal (2026), ensuring that variable-duration downstream tasks such as vector polygon dissolving never delay the next hydraulic warm start. A targeted fault-injection test confirmed that an injected failure in the downstream routing workflow leaves the simulation DAG's state file untouched and allows consecutive warm starts to proceed uninterrupted. While this study did not compare stateful simulation chaining against a precomputed library of design-storm flood maps, chaining allows the digital twin to track arbitrary rainfall trajectories without pre-selecting a matching storm template.

\subsection{Spatial Granularity and Performance Tradeoffs}

The 277-fold lookup speedup of $256 \times 256$ GiST-indexed PostGIS tiles over monolithic raster rows reflects the unit of database work, matching the benchmark findings of Agosto (2013): answering a coordinate query requires decoding only a single $256\text{ m} \times 256\text{ m}$ block rather than unpacking 55 million cells. Conversely, offloading Web Mercator vector tile pyramids (Li \& Li, 2024; Stefanakis, 2017) to MinIO object storage (Wan et al., 2016) isolates high-concurrency map browsing from PostgreSQL entirely, as confirmed by zero database query growth under a 50-client load test. Alongside these two tracks, computing a single road-edge depth table that drives the standing-flood continuation rule, the historical archive, and the Contract B binary export guarantees that all consumers evaluate the exact same maximum edge depths, while TimescaleDB temporal chunk compression (Zimányi et al., 2020) keeps long-term edge-depth archiving compact as consecutive runs accumulate.

\subsection{Operational Risk Management and System Resilience}

Because the underlying surface-only 3Di model lacks surveyed subsurface drainage and local rain gauges, its simulated depths represent physical hazard indicators rather than exact street measurements (Bui \& Tran, 2026). Contract B manages this operational risk structurally by separating hazard estimation from downstream decision policy: the pipeline publishes maximum edge water depths and run timestamps, whereas vehicle-specific closure limits, precautionary buffers, and route penalties reside entirely in the downstream routing consumer evaluated in the companion study (Tran et al., 2026). This boundary allows municipal operators to adjust vehicle safety margins or swap the upstream hydrodynamic solver without modifying the orchestration or serving tiers.

During pipeline development, the primary operational hazard was a silent data failure in which a coordinate-projection mismatch caused the polygon area filter to drop every feature and publish zero vector tiles while still returning a zero exit code. Consequently, geospatial digital twins must enforce semantic output checks, such as asserting non-zero feature and byte counts at each stage boundary, alongside task exit codes, as summarized in the degraded-mode behavior matrix (Table 3).

\subsection{Architectural Generalizability Across Urban Environmental Hazards}

Although demonstrated on pluvial flooding in Hanoi, the five structural components of the architecture (a threshold gate on the forecast feed, stateful cycle-to-cycle simulation chaining, GiST-indexed raster tiling, object-hosted vector tile pyramids, and a fixed-schema network impedance contract) apply broadly to urban hazards governed by gridded environmental fields, such as urban heat exposure or air-quality dispersion. Mandal et al. (2026) identified the latency gap between heavy physical solvers and rapid decision services as a central bottleneck of geospatial digital twins, and publishing a compact binary edge contract alongside decoupled serving tracks offers a reusable pattern to bridge that gap.

\begin{table}[!htbp]
\centering
\small
\setlength{\tabcolsep}{6pt}
\caption*{\textbf{Table 3}\\[0.2em]\emph{Degraded Conditions of the Pipeline, the Intended Behavior, and the Evidence}}
\begin{tabular}{>{\raggedright\arraybackslash}p{0.22\textwidth}>{\raggedright\arraybackslash}p{0.39\textwidth}>{\raggedright\arraybackslash}p{0.31\textwidth}}
\toprule
\textbf{Condition} & \textbf{Intended behavior} & \textbf{Evidence} \\
\midrule
Cloud solver unavailable & No new run is published, and the downstream stages do not start & Not tested \\
Weather feed unavailable & Not defined in this study & Not tested \\
Routing adapter unavailable & The simulation state and the map tiles are not affected, and routes are not updated & Confirmed by fault injection: the simulation state and warm starts continued uninterrupted \\
Depths out of date & A run whose slot is older than 120 minutes is skipped, so it cannot replace a fresh forecast, and the interface shows the time of the forecast step & Follows from the code. With no raster for the current hour, the depth chart stayed empty while the pipeline was paused \\
\bottomrule
\end{tabular}
\end{table}

\section{Conclusion}

\subsection{Summary of Contributions}

This paper set out to build and test an end-to-end prototype of an automated urban flood digital twin for Hanoi under data scarcity, turning a rainfall input into road-edge depths within a 15-minute cycle. A storm replay showed that a commodity server running around a cloud solver can meet this timing, as every run finished inside the cycle without overlapping consecutive steps. The first contribution, a pipeline of decoupled, event-driven workflows, kept downstream processing away from the simulation path and isolated the simulation state from downstream failures, as confirmed by a fault-injection check showing that warm starts continue uninterrupted. Building on this separation, the second contribution chained weather-gated runs through hydraulic restart states, keeping simulated floodwaters continuous across cycles instead of restarting from dry ground.

The third contribution, a dual-track spatial serving design, separated rapid point lookups on tiled storage from static map tile requests, enabling interactive dashboards and queries to share each forecast without database contention or simulation load. Finally, the fourth contribution established standardized output contracts with stable road identifiers, supplying a depth table for every road edge that allowed routing engines to enforce vehicle-specific travel limits. Together, these four contributions demonstrate that loosely coupled workflows, continuous restart states, and stable contracts can deliver actionable, edge-level flood guidance within an operational forecast cycle.

\subsection{Practical Implications for Scaling to Larger Areas}

Scaling this 15-minute cycle from the 32.8 km\textsuperscript{2} district domain to the whole of Hanoi remains an untested hypothesis that depends on how the local post-processing stages (raster extraction, vector tile generation, and road-edge sampling) grow with grid size alongside cloud solver quotas. Nevertheless, the formal separation between Contract A and Contract B provides a modular foundation for city-scale expansion: municipal agencies can update downstream routing policies, attach emergency dispatch or pump-scheduling consumers to the same published tables, or adapt the gated multi-DAG pipeline to other gridded urban hazards without redesigning the core orchestration engine.

\subsection{Limitations and Future Research Directions}

Several limitations define the scope of these findings and point to future research. First, the underlying hydrodynamic model lacks surveyed storm drainage, pumping stations, and local rain gauges, and the replay used half-hourly satellite precipitation estimates (53.88~mm in the 12-hour replay window) rather than live operational forecasts or the unwindowed 177.3 to 199~mm station totals reported for that night. Consequently, the simulated depths and the persistence of standing water after rainfall cessation reflect an undrained surface model and were calibrated against only eight newspaper points rather than validated against street-level sensors, nor were chained runs benchmarked against cold-started runs for depth accuracy. Second, all timing and storage measurements come from a single storm replay on one server without recording internal solver sub-steps, peak worker memory, or multi-week disk growth, and while downstream fault isolation was verified by fault injection, upstream cloud-solver or weather-feed outages and the pending display mode remain untested.

Finally, because the pipeline assimilates no real-time water-level observations and does not feed diverted traffic back into road impedances, it operates as an open-loop forecasting prototype rather than a closed-loop cyber-physical twin, and detailed evaluation of the vehicle-specific routing engine is reserved for a companion paper (Tran et al., 2026). Future work should therefore validate simulated depths against street-level gauge networks, stress-test the four-DAG pipeline on larger metropolitan domains and live forecast feeds, verify the standing-flood termination clause on a drained hydraulic model, and close the operational loop through real-time sensor assimilation.

\FloatBarrier

\clearpage

\section*{Code and Data Availability}

The pipeline source code, workflow definitions, spatial serving scripts, and output contract implementations are publicly available at \url{https://github.com/alice182012-dev/t3-digital-twin}. The 3Di hydrodynamic model and its proprietary inputs are owned by TECOTEC Technologies Joint Stock Company and are not publicly released.

\section*{Funding}

This research was funded by TECOTEC Technologies Joint Stock Company.

\clearpage

\section*{References}

\begingroup\small\setlength{\parindent}{0pt}\setlength{\parskip}{0.5em}\raggedright

\hangindent=1.5em\hangafter=1 Agosto, E. (2013). Vector--raster server-side analysis: A PostGIS benchmark. \emph{Applied Geomatics}, \emph{5}(2), 149--160. \url{https://doi.org/10.1007/s12518-013-0104-x}

\hangindent=1.5em\hangafter=1 Apel, H., Benisch, J., Helm, B., Vorogushyn, S., \& Merz, B. (2024). Fast urban inundation simulation with RIM2D for flood risk assessment and forecasting. \emph{Frontiers in Water}, \emph{6}, Article 1310182. \url{https://doi.org/10.3389/frwa.2024.1310182}

\hangindent=1.5em\hangafter=1 Bast, H., Delling, D., Goldberg, A., Müller-Hannemann, M., Pajor, T., Sanders, P., Wagner, D., \& Werneck, R. F. (2016). Route planning in transportation networks. In L. Kliemann \& P. Sanders (Eds.), \emph{Algorithm engineering: Selected results and surveys} (pp. 19--80). Springer. \url{https://doi.org/10.1007/978-3-319-49487-6_2}

\hangindent=1.5em\hangafter=1 Bui, M. H., \& Tran, M. A. (2026). \emph{Building and calibrating a 3Di subgrid flood model for Mỹ Đình and Tây Mỗ, Hanoi, without a surveyed terrain model, drainage data, or rain gauges} [Preprint]. arXiv. \url{https://arxiv.org/abs/2610.20789}

\hangindent=1.5em\hangafter=1 Caíno-Lores, S., Lapin, A., Carretero, J., \& Kropf, P. (2020). Applying big data paradigms to a large scale scientific workflow: Lessons learned and future directions. \emph{Future Generation Computer Systems}, \emph{107}, 1066--1083. \url{https://doi.org/10.1016/j.future.2018.04.014}

\hangindent=1.5em\hangafter=1 Casulli, V., \& Stelling, G. S. (2011). Semi-implicit subgrid modelling of three-dimensional free-surface flows. \emph{International Journal for Numerical Methods in Fluids}, \emph{67}(4), 441--449. \url{https://doi.org/10.1002/fld.2361}

\hangindent=1.5em\hangafter=1 Hanoi Sewerage and Drainage Company. (2024). \emph{Báo cáo công tác quản lý vận hành hệ thống thoát nước và phương án ứng phó úng ngập mùa mưa} [Report on the management and operation of the drainage system and the response plan for flooding in the rainy season]. Hanoi Sewerage and Drainage Company.

\hangindent=1.5em\hangafter=1 Huffman, G. J., Stocker, E. F., Bolvin, D. T., Nelkin, E. J., \& Tan, J. (2023). \emph{GPM IMERG Late Precipitation L3 Half Hourly 0.1 degree x 0.1 degree V07} [Data set]. NASA Goddard Earth Sciences Data and Information Services Center (GES DISC). \url{https://doi.org/10.5067/GPM/IMERG/3B-HH-L/07}

\hangindent=1.5em\hangafter=1 Kaynak, S., Kaynak, B., Mermer, O., \& Demir, I. (2025). City-scale digital twin framework for flood impact analysis: Integrating urban infrastructure and real-time data analytics. \emph{Urban Climate}, \emph{59}, Article 102640. \url{https://doi.org/10.1016/j.uclim.2025.102640}

\hangindent=1.5em\hangafter=1 Kefi, M., Mishra, B. N., Kumar, P., Masago, Y., \& Fukushi, K. (2018). Assessment of tangible direct flood damage using a spatial analysis approach under the effects of climate change: Case study in an urban watershed in Hanoi, Vietnam. \emph{ISPRS International Journal of Geo-Information}, \emph{7}(1), Article 29. \url{https://doi.org/10.3390/ijgi7010029}

\hangindent=1.5em\hangafter=1 Kim, Y., Oh, J., \& Bartos, M. (2025). Stormwater digital twin with online quality control detects urban flood hazards under uncertainty. \emph{Sustainable Cities and Society}, \emph{117}, Article 105982. \url{https://doi.org/10.1016/j.scs.2024.105982}

\hangindent=1.5em\hangafter=1 Li, G., \& Li, J. (2024). An adaptive-size vector tile pyramid construction method considering spatial data distribution density characteristics. \emph{Computers \& Geosciences}, \emph{186}, Article 105537. \url{https://doi.org/10.1016/j.cageo.2024.105537}

\hangindent=1.5em\hangafter=1 Luo, P., Mu, D., Xue, H., Ngo-Duc, T., Dang-Dinh, K., Takara, K., Nover, D., \& Schladow, G. (2018). Flood inundation assessment for the Hanoi Central Area, Vietnam under historical and extreme rainfall conditions. \emph{Scientific Reports}, \emph{8}(1), Article 12623. \url{https://doi.org/10.1038/s41598-018-30024-5}

\hangindent=1.5em\hangafter=1 Mandal, D., Zou, L., Wadhwa, A., Wilkho, R. S., Cai, Z., Zhou, B., Ye, X., Newman, G., Gharaibeh, N., \& Güneralp, B. (2026). FlowsDT: A geospatial digital twin for navigating urban flood dynamics. \emph{Computers, Environment and Urban Systems}, \emph{123}, Article 102414. \url{https://doi.org/10.1016/j.compenvurbsys.2026.102414}

\hangindent=1.5em\hangafter=1 Nagpal, M. (2026). A reproducible benchmark harness for comparing workflow orchestration engines: Temporal, Conductor, and Apache Airflow. \emph{SoftwareX}, \emph{33}, Article 102998. \url{https://doi.org/10.1016/j.softx.2026.102998}

\hangindent=1.5em\hangafter=1 Nguyen, H. D., Doan, T. B., Pham, S. D., \& Tran, D. D. (2026). Urban expansion trends and their relationship with flood susceptibility during the period 2014--2024 in Hanoi City. \emph{Vietnam Journal of Earth Sciences}, \emph{48}(2), 239--258. \url{https://doi.org/10.15625/2615-9783/24214}

\hangindent=1.5em\hangafter=1 Nhân Dân. (2026, September 17). \emph{Hà Nội: Mưa lớn nhất trong 3 ngày, hàng chục điểm ngập khắp thành phố} [Hanoi: The heaviest rain in 3 days, dozens of flooded spots across the city]. \url{https://nhandan.vn/ha-noi-mua-lon-nhat-trong-3-ngay-hang-chuc-diem-ngap-khap-thanh-pho-post989028.html}

\hangindent=1.5em\hangafter=1 Ruangpan, L., Mahgoub, M., Abebe, Y. A., Vojinovic, Z., Boonya-Aroonnet, S., Sanchez Torres, A., \& Weesakul, S. (2023). Real time control of nature-based solutions: Towards smart solutions and digital twins in Rangsit Area, Thailand. \emph{Journal of Environmental Management}, \emph{344}, Article 118389. \url{https://doi.org/10.1016/j.jenvman.2023.118389}

\hangindent=1.5em\hangafter=1 Stefanakis, E. (2017). Web Mercator and raster tile maps: Two cornerstones of online map service providers. \emph{Geomatica}, \emph{71}(2), 100--108. \url{https://doi.org/10.5623/cig2017-203}

\hangindent=1.5em\hangafter=1 Stelling, G. S. (2012). Quadtree flood simulations with sub-grid digital elevation models. \emph{Proceedings of the Institution of Civil Engineers - Water Management}, \emph{165}(10), 567--580. \url{https://doi.org/10.1680/wama.12.00018}

\hangindent=1.5em\hangafter=1 Tiền Phong. (2025a, October 7). \emph{Hà Nội còn trên 100 điểm ngập, có 23 điểm không thể lưu thông} [Hanoi still has more than 100 flooded spots, 23 of them impassable]. \url{https://tienphong.vn/ha-noi-con-tren-100-diem-ngap-co-23-diem-khong-the-luu-thong-post1784918.tpo}

\hangindent=1.5em\hangafter=1 Tiền Phong. (2025b, November 7). \emph{Vì sao trạm bơm tiêu lớn nhất Hà Nội vẫn chưa hoạt động hết công suất?} [Why is the largest drainage pumping station in Hanoi still not running at full capacity?]. \url{https://tienphong.vn/vi-sao-tram-bom-tieu-lon-nhat-ha-noi-van-chua-hoat-dong-het-cong-suat-post1793677.tpo}

\hangindent=1.5em\hangafter=1 Tran, M. A., Nguyen, V., Nguyen, V., \& Bui, M. H. (2026). \emph{Time-aware urban flood routing with sticky edge closure on forecast depth: A Hanoi case study} [Manuscript in preparation]. TECOTEC Technologies Joint Stock Company.

\hangindent=1.5em\hangafter=1 Tuổi Trẻ. (2025, October 7). \emph{Mưa xuyên đêm, nhiều tuyến phố Hà Nội thành sông} [Overnight rain turns many Hanoi streets into rivers]. \url{https://tuoitre.vn/mua-xuyen-dem-duong-pho-ha-noi-bat-dau-ngap-sau-20251007070931343.htm}

\hangindent=1.5em\hangafter=1 Tuổi Trẻ. (2026, July 9). \emph{Mưa 30 phút, một số tuyến đường Hà Nội đã ngập} [Thirty minutes of rain, some Hanoi streets already flooded]. \url{https://tuoitre.vn/mua-30-phut-mot-so-tuyen-duong-da-ngap-ha-noi-kiem-tra-he-thong-ngap-vua-xay-dung-100260709181057919.htm}

\hangindent=1.5em\hangafter=1 van Dam, A., Kleptsova, O., Stelling, G., \& Pleumeekers, O. (2012). 3Di subgrid flooding. In F. Klijn \& T. Schweckendiek (Eds.), \emph{Comprehensive flood risk management} (pp. 519--523). CRC Press. \url{https://doi.org/10.1201/b13715-72}

\hangindent=1.5em\hangafter=1 Wan, L., Huang, Z., \& Peng, X. (2016). An effective NoSQL-based vector map tile management approach. \emph{ISPRS International Journal of Geo-Information}, \emph{5}(11), Article 215. \url{https://doi.org/10.3390/ijgi5110215}

\hangindent=1.5em\hangafter=1 Xie, T., Zhang, M., Lian, J., Shen, H., \& Wen, J. (2026). A generalizable WebGIS--machine learning system for real-time urban flood forecasting. \emph{Urban Climate}, \emph{65}, Article 102741. \url{https://doi.org/10.1016/j.uclim.2025.102741}

\hangindent=1.5em\hangafter=1 Zimányi, E., Sakr, M., \& Lesuisse, A. (2020). MobilityDB: A geospatial-temporal database system. \emph{ACM Transactions on Database Systems}, \emph{45}(4), 1--42. \url{https://doi.org/10.1145/3406534}

\endgroup

\end{document}